\documentclass[11pt,twoside]{article}

\usepackage{asp2006}
\usepackage{epsf}
\usepackage{graphicx}
\usepackage{lscape}

\newcommand{\spitzer}{{\it Spitzer}}
\newcommand{\sst}{{\it Spitzer Space Telescope}}

\sfcode`A=1000 \sfcode`B=1000 \sfcode`C=1000 \sfcode`D=1000
\sfcode`E=1000 \sfcode`F=1000 \sfcode`G=1000 \sfcode`H=1000
\sfcode`I=1000 \sfcode`J=1000 \sfcode`K=1000 \sfcode`L=1000
\sfcode`M=1000 \sfcode`N=1000 \sfcode`O=1000 \sfcode`P=1000
\sfcode`Q=1000 \sfcode`R=1000 \sfcode`S=1000 \sfcode`T=1000
\sfcode`U=1000 \sfcode`V=1000 \sfcode`W=1000 \sfcode`X=1000
\sfcode`Y=1000 \sfcode`Z=1000

\begin{document}
\title{\spitzer/IRAC Characterization of Galactic AGB Stars}
\author{Massimo Marengo, J.\ L.\ Hora, P.\ Barmby, S.\ P.\
  Willner, Lori E.\ Allen, Michael T.\ Schuster and G.\ G.\ Fazio}
\affil{Harvard-Smithsonian CfA, Cambridge, MA 02138, USA}

\begin{abstract} %%% Abstract to run on from here.
The \sst\ and in particular its InfraRed Array
Camera (IRAC) is an ideal facility to study the distribution of AGB
stars in our own and other galaxies because of its efficiency in
surveying vast areas of the sky and its ability to
detect sources with infrared excess. The IRAC colors of  AGB stars,
however, are not well known because cool stars have numerous
molecular absorption
features in the spectral region covered by the IRAC photometric
system. The presence and strength of these features depends on the
chemistry of the stellar atmosphere and the mass loss rate and can change
with time due to the star's variability. To characterize
the IRAC colors of AGB stars, we are carrying out a \spitzer{}
Guaranteed Time Observation program to observe a sample of AGB
stars with IRAC. The results will be made available to the community
in the form of template magnitudes and colors for each target with
the goal of aiding the identification of AGB stars in already
available and future IRAC surveys. We present here the first results
of this project.
\end{abstract}

%%% MAIN BODY OF TEXT GOES HERE. CONSULT "INSTRUCTIONS FOR AUTHORS USING
%%% LATEX2E MARKUP", SECTIONS 2.3-2.6 FOR HELP WITH EQUATIONS, FIGURES,
%%% AND TABLES.

%\section{}   %%% Top level section head (remove "%" symbol)
%\subsection{}   %%% Second level section head (remove "%" symbol)
%\subsubsection{}   %%% Lowest level section head (remove "%" symbol)
%\section*{}    %%% Unnumbered top level section head (remove "%" symbol)
%\subsection*{}   %%% Unnumbered second level section head (remove "%" symbol)

\section{Mapping AGB Stars with \spitzer/IRAC}

Asymptotic Giant Branch (AGB) stars, with luminosities exceeding
$10^4$~$L_\odot$, are among the brightest stars in the galaxy. They are
also among the reddest because of intense mass loss processes (up to
$10^{-4}$~$M_\odot$/yr) responsible for enshrouding AGB stars in dusty
envelopes, which are the source of strong infrared excess. These
characteristics of AGB stars make them very important tools
to study the structure of our and other galaxies by mapping their
distribution at infrared wavelengths. The \sst\ 
\citep{werner04} is especially suited for this task.  The telescope
has very low background emission, which makes the InfraRed Array Camera (IRAC,
\citealt{fazio04}) onboard \spitzer\ very
sensitive and allows  mapping large areas of the sky in a very short
time. Several large area IRAC surveys, including GLIMPSE
\citep{benjamin03} covering 220 square degrees of the galactic plane
and SAGE  \citep{meixner06}, mapping a 50 square degrees area centered
on the Large Magellanic Cloud,
are already available.  The four IRAC
channels operating at 3.6, 4.5, 5.8, and 8.0~\micron{} allow detection
of both photospheric and dust shell emission.  The IRAC colors of AGB stars,
however, are not very well known.

AGB stars have
numerous molecular absorption features in the spectral region covered
by the IRAC photometric system \citep{waters99}, as demonstrated by
observations with the Short Wavelength Spectrometer
\citep[SWS,][]{valentijn96} aboard
the {\it Infrared Space Observatory} (ISO).  Features seen include 
H$_2$O, SiO, CO$_2$, CO, and silicates
in stars with atmospheric C/O ratio $<$1, while C$_2$H$_2$, HCN, CS,
C$_3$, and carbonaceous dust (SiC and amorphous carbon) have been found
in carbon stars, which have ${\rm C/O} > 1$. The presence and strength of
these spectral features depends on the chemistry of the stellar
atmosphere \citep{sloan98, sloan98b} and the mass loss rate. The features can
also change with time \citep{onaka02} because AGB stars are long
period variables of Mira, semi-regular, or irregular type.

As a result of these dust and molecular features, the IRAC colors of
AGB stars can be quite different from the ``reddened photospheres''
that one would expect for mass losing giants of late spectral
type. There is a need, in order to efficiently identify AGB
stars among other red objects, for accurate measurements of IRAC
colors of AGB stars. For this reason, we have started a program to
observe a sample of nearby AGB stars with IRAC as
part of Cycle-3 Guaranteed Time Observations (GTO). The program is
in progress, and we present here the first results.

\begin{figure}
\begin{center}
\includegraphics[angle=0,width=0.9\textwidth]{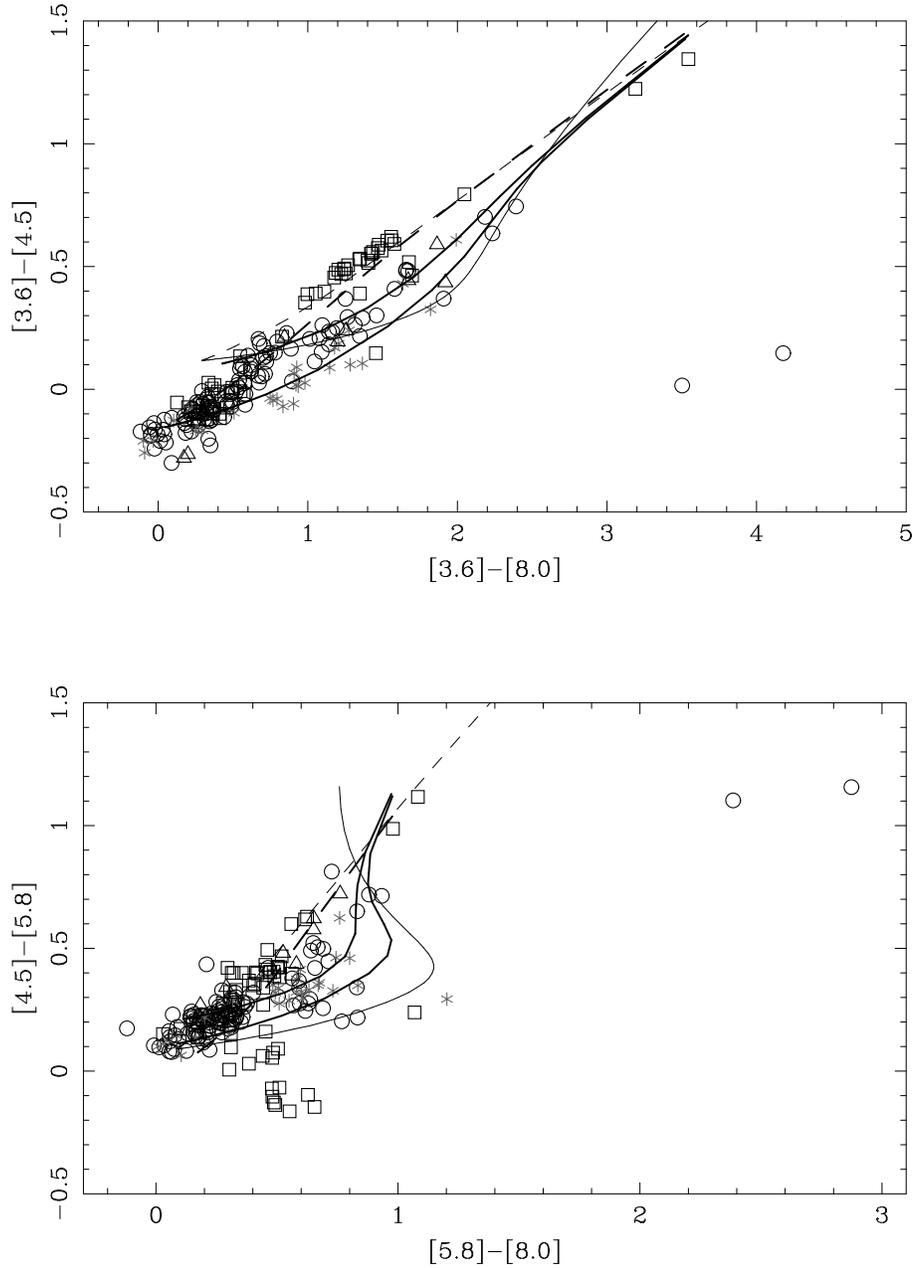}
\end{center}
\caption{IRAC synthetic color-color diagrams of O-rich AGB stars
  (circles), carbon stars (squares), S-stars (triangles), and M-type
  supergiants (asterisks) derived from ISO SWS spectra. The tracks are
  models of circumstellar envelopes with silicate (solid lines) and
  amorphous carbon (dashed lines) dust. Thin lines are DUSTY models
  using a black body as central source, and the thick lines are models
  computed by \citet{groenewegen06} using realistic stellar atmosphere
  models. The two O-rich outliers are two-epoch measurements of V354
  Lac, an AGB star with a very
  thick envelope.}
\end{figure}

\section{IRAC AGB Colors from ISO SWS and Models}

As a first step in deriving IRAC colors for AGB stars, we 
computed synthetic colors for 87 O-rich AGB stars, 27 carbon
stars, 8 S-stars, and 23 M-type supergiants
\citep[lists from][]{sloan98b,sloan98}
by convolving their ISO SWS spectra with the IRAC
bandpasses. We obtained the ISO spectra (212 in total, with some
stars observed in multiple epochs) from \citet{sloan03} and the IRAC
transmission profile from the \spitzer{} Science Center web
site\footnotemark{}.

\footnotetext{http://ssc.spitzer.caltech.edu/irac/spectral\_response.html}

The distribution of the sources on two IRAC color-color diagrams is
shown in Figure~1:

\begin{enumerate}
\item Sources with larger mass loss, as expected, have a larger
  infrared excess, especially in the $[3.6]-[8.0]$ color that is the
  most sensitive to the overall slope of the spectrum.
\item IRAC colors are not sensitive to the detailed dust composition
  because even the longest passband, centered around 8.0~\micron{},
  only marginally includes the 10~\micron{} silicate feature.
\item Carbon stars tend to have redder $[3.6]-[4.5]$ colors than O-rich
  stars having the same infrared excess. This is due to very strong
  C$_2$H$_2$ and HCN absorption within the 3.6~\micron{} band in
  carbon stars and broad CO$_2$ and SiO absorption features in the
  4.5~\micron{} band in O-rich stars.
\item There is a population of carbon stars with very blue ($-$0.5
  magnitudes below the model tracks) $[4.5]-[5.8]$ color due to strong
  C$_3$ absorption at 5.8~\micron. This feature is transient, as
  some sources observed by ISO at multiple epochs have this
  anomalous color only at some pulsation phases.
\item S-stars and M-type supergiants tend to have colors very
  similar to the colors of M-type AGB stars.
\end{enumerate}

Figure~1 also shows the tracks of simple radiative transfer models in
the IRAC colors. The DUSTY\footnotemark{} models, which  use a
cool black body for the spectrum of the central source, cannot fit
the colors of sources with little
or no infrared excess because the models  are missing the molecular features
responsible for blue IRAC colors. The  models
computed by \citet{groenewegen06} using realistic spectra for the
central AGB star fit the data much better. 
On the whole, the model tracks intercept the synthetic colors
for most of the sources with the exception of the carbon stars
that show  strong C$_3$ absorption features. 

\footnotetext{http://www.pa.uky.edu/$\sim$moshe/dusty}

\section{The \spitzer/IRAC GTO Program}

The nearby AGB stars selected for IRAC observations are ones for
which accurate parallax or interferometric distances
and reliable determinations of their mass loss rates are available. The
observations target 22 O-rich stars, 7 intrinsic S-stars, 19
carbon stars, and 4 M-type supergiants. The supergiants have been added
as a comparison sample of mass-losing stars outside the AGB. The mass
loss rates of our target stars range from $10^{-8}$ to
$10^{-4}$~$M_\odot/{\rm yr}$ in each category \citep{guandalini06}. The
sample contains Mira, semi-regular, and irregular variables. 

The goal of the program is to measure accurate IRAC photometry for all
stars in the sample in order to validate and cross-calibrate the ISO SWS
spectra for this class of sources. This is particularly important
because the absolute photometric calibration of IRAC is based mainly
on A-type primary calibrators
\citep{reach05} and because of the uncertainties in the ISO SWS
absolute calibration resulting from the splicing of the different
spectral segments. Each star will be observed in two epochs two months
apart in order to evaluate the change in color at different
variability phases.  (The timing of the observations is constrained
by the orbit of the spacecraft).

\begin{figure}
\begin{center}
\includegraphics[angle=0,width=0.85\textwidth]{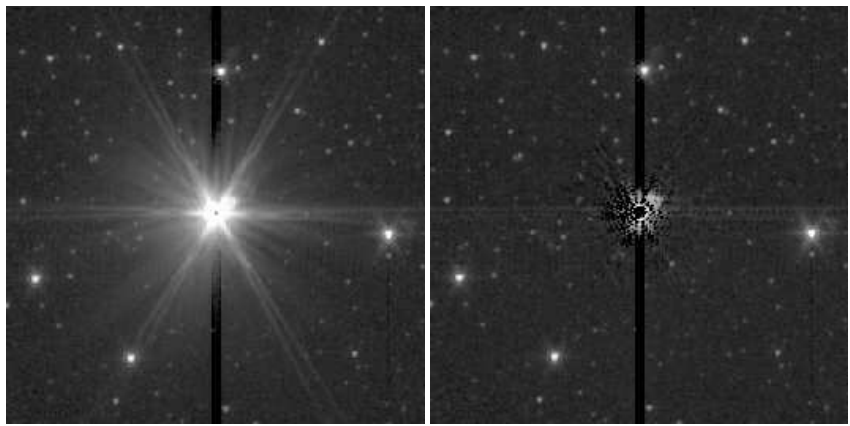}
\end{center}
\caption{IRAC 4.5~\micron{} image of SZ~Car (SRb carbon star with
  $m_{[4.5]} = 1.18$). Left panel is the mosaic resulting from the
  co-addition of 5 individual 2-second frame time exposures. The
  saturated star occupies almost all the IRAC field of view
  (5$\times$5 arcmin). The right panel shows the same image after PSF
  subtraction. The PSF fitting determines the magnitude of the AGB
  star with an accuracy of 1 -- 3\% in flux.}
\end{figure}

Nearby AGB stars are of course among the brightest objects in the
infrared sky. This makes the IRAC photometry very
difficult because all sources will be heavily saturated even with the
shortest available frame times. To solve this problem, we have
developed a technique to derive the Vega magnitudes of saturated stars
by fitting the low level features of the saturated PSF (diffraction
spikes and PSF wings) with a model of the IRAC PSF derived from a
sample of bright stars \citep{marengo06}. Figure~2 shows the
4.5~\micron{} image of SZ~Car (SRb carbon star), one of the first objects
observed in our program, before and after PSF subtraction. The
stability of the \spitzer{} optical system allowed us to create a very
accurate model of the IRAC PSF.  The PSF is directly normalized to
the actual image of Vega, 
which is one of the stars used in its construction. By fitting the observed
saturated sources with this PSF, we can derive their 
magnitudes with an accuracy within 1 -- 3\% independently of the
standard IRAC flux calibration \citep{schuster06}.

The program is in progress. As of 2006 November, 33 stars have been
observed in their first epoch. Preliminary results show that the photometry is
consistent with the synthetic colors derived from ISO SWS. The program
will be completed within one year, after which we will release the
complete catalog to the community.

\acknowledgements %%% Text of acknowledgements runs on after this command.
This work is based in part on observations made with the \sst,
which is operated by the Jet Propulsion Laboratory, California 
Institute of Technology under a contract with NASA. Support for this work 
was provided by NASA through an award issued by JPL/Caltech.

%%% THE BIBLIOGRAPHY
%%%
%%% CONSULT SECTION 3 OF "INSTRUCTIONS FOR AUTHORS" FOR HOW TO USE NATBIB.
%%% AUTHORS ARE ENCOURAGED TO USE EITHER THE "THEBIBLIOGRAPY" ENVIRONMENT
%%% BY UNCOMMENTING (DELETING THE "%" SYMBOL) THE COMMANDS BELOW, OR BY
%%% USING THE BIBTEX ENVIRONMENT. TO FIND OUT WHICH IS APPLICABLE TO YOUR
%%% CONTRIBUTION, CONSULT THE VOLUME EDITORS FOR YOUR PROCEEDINGS.
%%%

\end{document}